# RECTANGULAR AND CIRCULAR ANTENNA DESIGN ON THICK SUBSTRATE

Harsh Kumar & Shweta Srivastava

**Abstract** - Millimeter wave technology being an emerging area is still very undeveloped. A substantial research needs to be done in this area as its applications are numerous. In the present endeavor, a rectangular patch antenna is designed on thick substrate and simulated using SONNET software, also a novel analysis technique is developed for circular patch antenna for millimeter wave frequency. The antenna is designed at 39 GHz on thick substrate and has been analyzed and simulated. The results of the theoretical analysis are in good agreement with the simulated results.

**Index Terms**— Microstrip, thick substrate, coaxial feeding.

——————————— ◆ ———————————

## 1 INTRODUCTION

MILLIMETER Wave can be classified as electromagnetic spectrum that spans between 30GHz to 300 GHz, which corresponds to wavelengths from 10mm to 1mm. Despite millimeter wave technology has been known for many decades, it is still undeveloped. The major applications of this technology being high speed point to point wireless local area network and broad band access[1]. The mm wave systems have mainly been deployed for military applications and low-cost integration solutions. It has started to gain a great deal of momentum from academia and industry due to its advantages limited access till date.

In this paper, however, the focus is specifically on 38 to 40 GHz frequency range, which is used for high speed microwave data link. Work has been done on rectangular antenna on thick substrate and empirical formula for input impedance has been also derived. Using this, a rectangular patch antenna is designed on thick substrate and simulate on SONNET software. The empirical formula for input impedance of the circular patch antenna on thick substrate is developed and a circular patch millimeter wave antenna is designed for 39 GHz on thick substrate. Comparison has been done between the simulated and the calculated results. When thick substrate is used surface wave excitation takes place so surface wave loss also has to be considered. James & Henderson [2] proposed that for thick substrate $h/\lambda_0 > 0.09$ For $\varepsilon_r$ =2.32 and $h/\lambda_0 > 0.03$ For $\varepsilon_r$ =10 where 'h' is the thickness of substrate and $\lambda_0$ is the free space wavelength.

————————————————

- *Harsh kumar is with the Department of ECE, Birla Institute of Technology, Mesra, Ranchi, India.*
- *Dr. Shweta Srivastava is with the Department of ECE, Birla Institute of Technology, Mesra, Ranchi, India.*

## 2 DESIGN OF RECTANGULAR PATCH MILLIMETER WAVE ANTENNA ON THICK SUBSTRATE

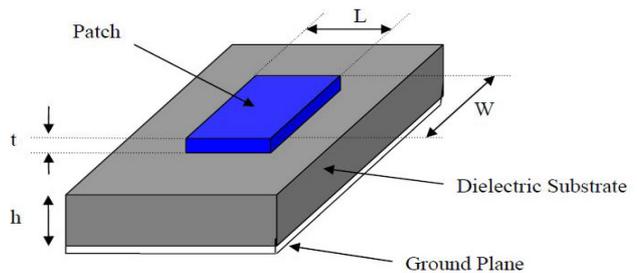

Fig. 1 Rectangular microstrip patch antenna

The formulas available for the design of thin substrate patch antenna are not applicable for thick substrate. The antenna which is designed from these formula is not resonate when antenna thickness is greater then $0.09\lambda_0$.

### 2.1 RECTANGULAR PATCH MILLIMETER WAVE ANTENNA

Patch resonant length and patch width were determined by analyzing the shape and slope of curve substrate thickness with experimental Length/calculated Length and substrate thickness with experimental Width / calculated Width respectively. The resulting formulas are [13]

$$L = \frac{\pi}{\epsilon_r\, r}\sqrt{h\lambda_d} \qquad (1)$$

$$W = \frac{\lambda_d \varepsilon_{rl}}{2\pi}\left[\ln\left(\frac{\lambda_d}{h}\right) - 1\right] \quad (2)$$

$$\varepsilon_{rl} = 0.5\left[(\varepsilon_r + 1) + (\varepsilon_r - 1)\left(1 + \left(\frac{10h}{L}\right)\right)^{-\frac{1}{2}}\right] \quad (3)$$

## 2.2 DETERMINATION OF INPUT RESISTANCE OF RECTANGULAR PATCH WITH THICK SUBSTRATE

A formula based on the cavity model and equivalent loss resonant circuit has been developed to calculate the resonant input resistance of the rectangular antenna elements with substrate thickness satisfied the criteria $h \geq 0.0815\lambda_0$. The antenna elements were consider in its fundamental mode and assumed infinite ground plane. The resonant input resistance of such antenna can be calculated from

$$R_{in} = R_c + R_d + R_r + R_s \quad (4)$$

The resistance due to conductor loss is calculated from

$$R_c = 0.00027 \frac{L}{W} Q_r^2 \sqrt{f_r(GHz)} \quad (5)$$

Where

$$Q_r = \frac{\lambda_0}{4h}\sqrt{\varepsilon_r} \quad (6)$$

Where $\varepsilon_{ew}$ is the effective permittivity of the substrate and is calculated from

$$\varepsilon_{rl} = 0.5\left[(\varepsilon_r + 1) + (\varepsilon_r - 1)\left(1 + \left(\frac{10h}{L}\right)\right)^{-\frac{1}{2}}\right] \quad (7)$$

The resistance due to dielectric loss is calculated from
The radiation resistance can be calculated from

$$R_r = \frac{U_r}{2\pi f c_r} \quad (8)$$

Where $L_{ef}$ is given by

$$L_{ef} = L + \left(\frac{W_{eq} - W}{2}\right)\left(\frac{\varepsilon_{ew} + 0.3}{\varepsilon_{ew} - 0.258}\right) \quad (9)$$

Where $W_{eq}$ is the equivalent patch width

$$W_{eq} = \frac{R_0 a}{Z_{cw}\sqrt{\varepsilon_{ew}}} \quad (10)$$

The existence of dielectric substrate over the conducting ground plane in microstip antenna can cause the surface wave excitation along air dielectric interface. The resistances due to surface wave excitation $R_s$ can be derive from ratio of power loss to surface wave $P_s$ and radiation power given by James

$$\frac{P_s}{P_r} = \frac{R_r}{R_s} = \frac{1}{T_1} \quad (11)$$

$$R_s = T_1 R_r \quad (12)$$

Where

$$K_1 = \sqrt{\frac{-\varepsilon_r^2 + \varepsilon_r\sqrt{\varepsilon_r^2 + 4K_0^2 h^2(\varepsilon_r - 1)}}{2h^2}} \quad (13)$$

$Z_{ew}$ is the characteristic impedance of the substrate filled strip line of width W and strip conductor of zero thickness. It is given for W/h=< 3.3 [13]

$$Z_{cw} = \frac{R_0}{\pi\sqrt{2(\varepsilon_r+1)}}\left[\ln\left(\frac{4h}{w} + \sqrt{2 + \frac{16h^2}{w^2}}\right) - \left(\frac{\varepsilon_r-1}{\varepsilon_r+1}\right)\left(0.2258 + \frac{0.1208}{\varepsilon_r}\right)\right] \quad (14)$$

For W/h>= 3.3, then

$$Z_{cw} = \frac{R_0}{2\sqrt{\varepsilon_r}}\left[\frac{W}{2h} + 0.4413 + \frac{0.0823(\varepsilon_r-1)}{\varepsilon_r^2} + \left(\frac{\varepsilon_r-1}{\varepsilon_r}\right)\left(0.231 + 0.1592\ln\left(\frac{4h}{w} + 0.94\right)\right)\right]^{-1} \quad (15)$$

Where $R_0 = 120\pi$ is vacuum medium resistance. $Z_{0w}$ is the characteristic impedance of an air filled substrate characteristic impedance of an air filled substrate $\varepsilon_r = 1$.
Input impedance

$$R_{in} = R_0 \frac{L_{ef}}{\lambda_0\left(\varepsilon_{rl} - \frac{1}{2}\varepsilon_{rd}\right)}\left\{1 - \frac{\sin[2k_0(a+\Delta L)]}{[2k_0(a+\Delta L)]}\right\}\{1 - \cos[2k_0(a+\Delta L)]\}^{-1} \quad (16)$$

Where

$$\Delta L = 0.412h\frac{(\varepsilon_{rl}+0.3)}{(\varepsilon_{rl}-0.258)}\left(\frac{L}{h} + 0.264\right)\left(\frac{L}{h} + 0.813\right)^{-1} \quad (17)$$

## 2.4 DESIGN SPECIFICATIONS

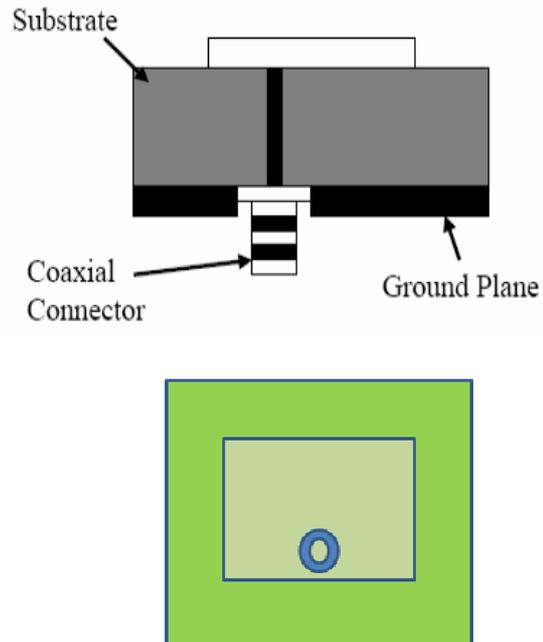

Fig 2: Rectangular patch with coaxial feeding

## 2.4 DESIGN DETAILS

- Frequency of operation (*fo*): The resonant frequency selected for this design is 39 GHz.
- Dielectric constant of the substrate (ε*r*):  ε*r*= 4.7
- Height of dielectric substrate (*h*):  h=0.8
- Length=1.06mm
- Width=0.98mm
- Feed point (a)=0.05mm

## 2.5 RESULTS

The theoretical results for return loss and VSWR were obtained using the above derived expressions (Fig.) and the antenna was simulated using sonnet software and the resulting graph for return loss is shown in Fig..

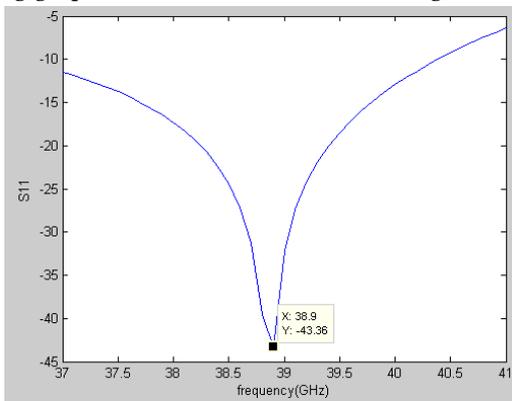

Fig.3 Theoretical results of variation of return loss with frequency

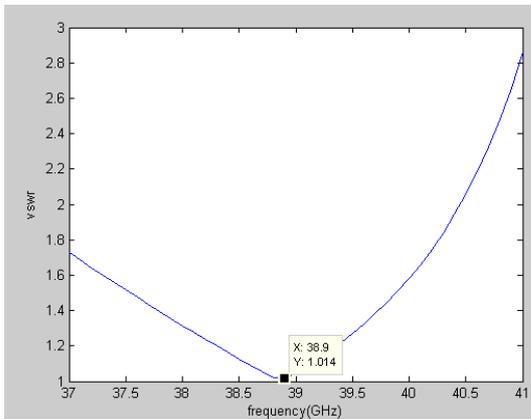

Fig.4: Theoretical results of variation of VSWR with frequency

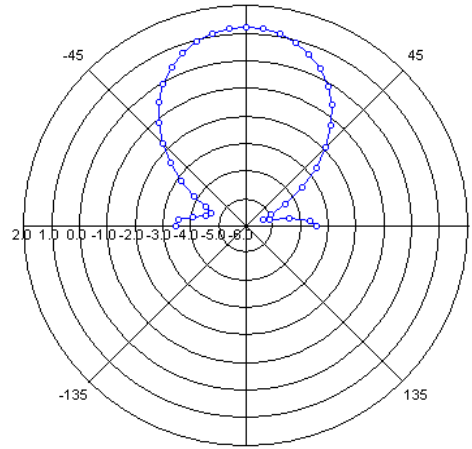

Fig 5: Radiation patern

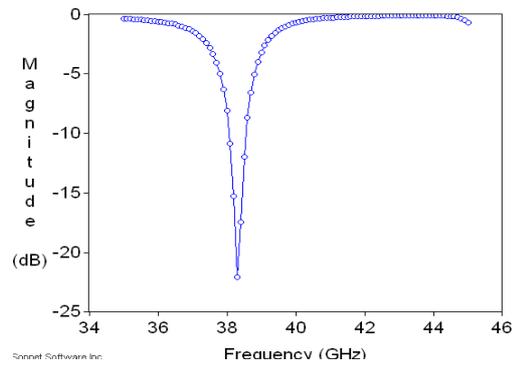

Fig 6: Simulated Return loss Vs frequency

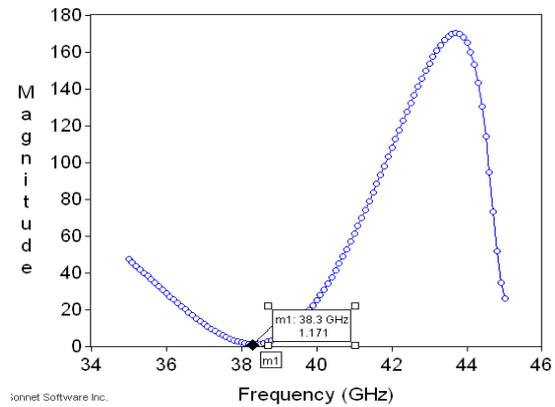

Fig 7: Simulated VSWR Vs frequency

In theoretical analysis resonance takes place at 38.9 GHz and the return loss is -41.36dB and  vswr is 1.014 but in simulation the resonance takes place at 38.3 GHz and return loss is-23.8dB,  vswr is1.171and gain of the antenna is 1.2dB

# 3 CIRCULAR PATCH ANTENNA

The expressions to find out the dielectric conductor radiation and losses in a microstrip antenna are given as follows [3, 4]:

$$P_r = \frac{1}{2\eta_0}\int_0^{2\pi}\int_0^{\pi/2}(|E_\theta|^2+|E_\varphi|^2)r^2\sin\theta\,d\theta\,d\varphi \quad (1)$$

$$P_c = 2\frac{R_s}{2}\iint |H_s|^2 dS = \frac{\omega W_T}{h\sqrt{\pi f \mu_0 \sigma}} \quad (2)$$

$$P_d = \frac{\omega \epsilon_0 \epsilon_r \tan\delta}{2}\iiint |E_z|^2 dV = \omega\tan\delta\, W_T \quad (3)$$

$P_d$ is dielectric loss, $P_c$ is conductor loss, $P_r$ is radiation loss, $\tan\delta$ loss tangent, $W_T$ is the total power absorbed, $\omega$ is the resonant frequency.

## 3.1 DETERMINATION OF INPUT IMPEDANCE FOR CIRCULAR PATCH ANTENNA ON THICK SUBSTRATE

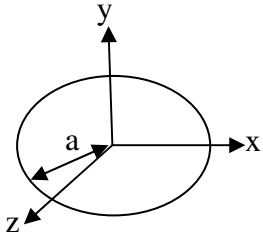

Fig. 8: Circular patch antenna

The relations for radiated power and radiation resistance for a circular patch antenna are given as [5]:

$$P_r = \frac{\pi^3 a^2 (E_0 h)^2}{2\lambda_0^2 \eta_0}\left[\frac{4}{3} - \frac{8}{15}(k_0 a)^2 + \frac{11}{105}(k_0 a)^4\right] \quad (4)$$

$$P_r = \frac{1}{2}G_r(E_0 h)^2$$

$$R_r = \frac{1}{\frac{\pi^3 a^2}{\lambda_0^2 \eta_0}\left[\frac{4}{3} - \frac{8}{15}(k_0 a)^2 + \frac{11}{105}(k_0 a)^4\right]} \quad (5)$$

where $R_r$ is Radiation resistance

The existence of dielectric substrate over the conducting ground plane in microstrip antenna can cause the surface wave excitation along air dielectric interface. The resistances due to surface wave excitation $R_s$ can be derive from ratio of power loss to surface wave $P_s$ and radiation power given by James[6]

$$\frac{P_s}{P_r} = \frac{R_r}{R_s} = \frac{1}{T_1}$$

$$R_s = T_1 R_r \quad (6)$$

where

$$T_1 = \left(\frac{K_1 h}{\epsilon_r}\right)^2\left[\left(1+\frac{1}{3}K_1^2 h^2\right)^2 + \left(1+\frac{1}{3}K_1^2 h^2\right)^2 \cos^{-2}(K_1 h)\right] \quad (7)$$

$$K_1 = \sqrt{\frac{-\epsilon_r^2 + \epsilon_r\sqrt{\epsilon_r^2 + 4K_0^2 h^2(\epsilon_r - 1)}}{2h^2}}$$

$W_T$ is the total energy stored in the patch at the resonance frequency for circular patch and is given by the expression

$$W_T = \frac{\epsilon_0 \epsilon_r h \pi E_0^2}{2}\int_0^a J_n^2(k\rho)\rho\,d\rho = \frac{E_0^2 h}{8\omega f \mu_0}J_n^2(k\rho)((k\rho)^2 - n^2) \quad (8)$$

Substituting the value of $W_T$ in eqn. 1 and 2 the value of power loss due to conductor and power loss due to dielectric for thick substrates is determined as

$$P_d = \omega\tan\delta\,\frac{E_0^2 h}{8\omega f \mu_0}J_n^2(k\rho)((k\rho)^2 - n^2) \quad (9)$$

$$P_c = \frac{\omega}{h\sqrt{\pi f \mu_0 \sigma}}\frac{E_0^2 h}{8\omega f \mu_0}J_n^2(k\rho)((k\rho)^2 - n^2) \quad (10)$$

Using these relations the value of resistance due to dielectric and conductor can be determined

$$R_d = \frac{4\mu_0 f h}{\tan\delta J_n^2(k\rho)((k\rho)^2 - n^2)} \quad (11)$$

$$R_c = \frac{4\mu_0 f h^2 \sqrt{\pi f \mu_0 \sigma}}{J_n^2(k\rho)((k\rho)^2 - n^2)} \quad (12)$$

Total resistance $R_T$ can be determine by using above equations

$$R_T = R_r + R_s + R_c + R_d \quad (13)$$

As given by Kara [7] the total resistance can be equated to input resistance of the microstrip antenna.

$$R_T = R_{in} \quad (14)$$

Using the above expression the reflection coefficient and return loss of the antenna can be determined.
Radiation efficiency can be defined as the ratio of radiated power to input power, that is

$$e_r = \frac{R_r}{R_r + R_s + R_c + R_d} \quad (15)$$

Directivity is defined as the ratio of the maximum power density in the main beam direction to the average radiated power density and can be expressed as [9]

$$D = \frac{\frac{r}{2\eta_0}(|E_\theta|^2 + |E_\varphi|^2)|_{\theta=0}}{P_r/4\pi} \quad (16)$$

$$G = e_r D \quad (17)$$

Fig.10 Theoretical and Simulated results of variation of return loss with frequency

### 3.2 Design consideration

For thick substrate the height of the substrate is decided by the given condition h=0.8 mm, $\varepsilon_r$ =2.32. The disk metallization radius 'a' can be determined by the resonance condition, that is, $J_n'(k_0 a\sqrt{\varepsilon_r})$ =0. For the lowest order mode n=1 and the 1st root of $J_n'$ occurs at 1.841[9]. The feed point $(\rho_0, \varphi_0)$ can be determined by the following expression [8].

$$R_{in} = R_r \frac{J_1^2(k_{11}\rho_0)}{J_1^2(k_{11}a)}$$

Using the above relations the dimensions of the patch were determined as radius, a=1.21 mm, feed point location ρ=0.46 for optimum matching

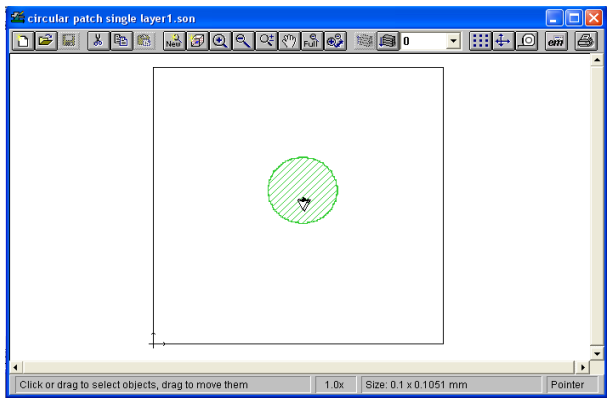

Fig.9:Block diagram of designed antenna (sonnet)

### 3.3 RESULTS

The theoretical results for return loss were obtained using the above derived expressions and the antenna was simulated using sonnet software and the resulting graph for return loss is shown in Fig.10.

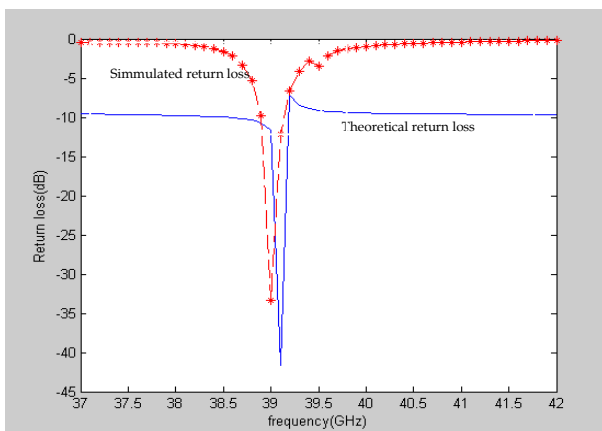

The graphs show a good agreement of the theoretical and simulated results, thus proving the correctness of the technique developed. Both the graphs are resonating at 39GHz. The bandwidth shown by the theoretical results (320 MHz) is lower than the simulated results (250 MHz). The VSWR value for theoretical analysis at reson ance comes out to be 1.38 and for simulated results it has the value of 1.3.

Gain is calculated using eqns 15, 16 and 17 and the results plotted in Fig. 5 showing the value for maximum gain obtained at resonance as 4.76dB

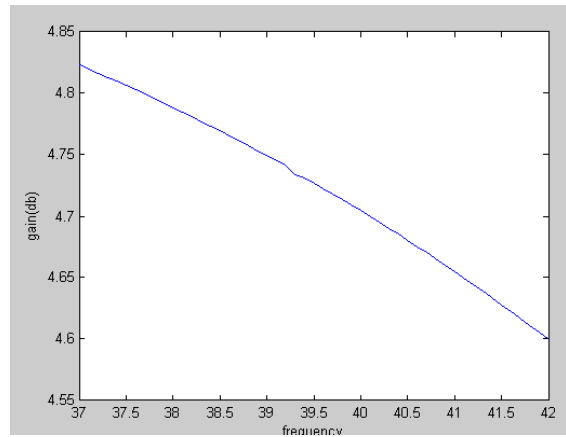

Fig.12:gain vs frequency(theoretical)

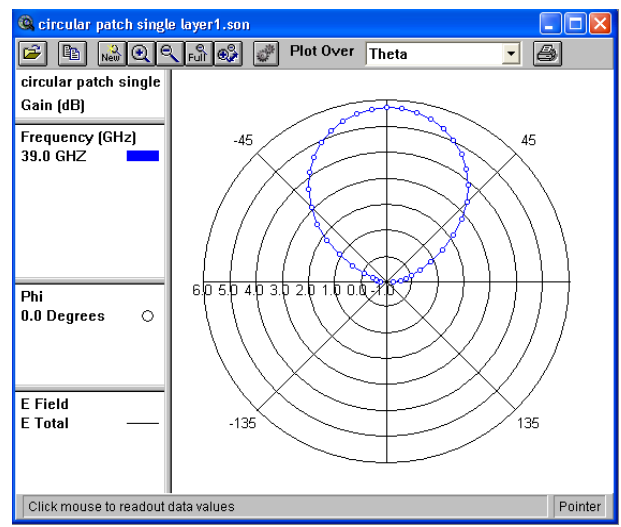

Fig.12: Simulate radiation patern

For simulated result we can see the gain of the antenna is 5.7dB at 39 GHz frequency

## 7 CONCLUSION

It can be concluded from the above analysis that an efficient technique has been developed for analysis of retangular and circular patch antenna at millimeter wave frequencies for thick substrates. The feasibility of the technique is proved by simulations. The expressions to find out input resistance and gain for circular patch antenna on thick sustrate have been developed and verified

## REFERENCES

[1] R. Piesiewicz, T. Kleine-Ostmann, N. Krumbholz, D. Mittleman, M. Koch, J. Schoebel, and T. Kurner,"Short-range ultra-broadband terahertz communications: concept and perspectives". *IEEE Antennas Propag*. Mag 49(6), 24–38 (2007)Dec.

[2] P. Kumar, A. K. Singh, G. Singh, T. Chakravarty and S. Bhooshan. "Terahertz technology – a new direction." Proc. *IEEE Int. Symp. Microwave*, pp. 195-201 (2006).

[3] Zhang Y P, Sun M and Guo L H 2005 On-chip antennas for 60-GHz radios in silicon technology *IEEE Trans. Electronic Devices* **52** 1664–8

[4] Mehmet kara , "Empirical formulas for the computation of physical parameter of rectangular microstrip antenna with thick substrate". *john wiley & son* ccc 0895-2477/97

[5] Zhang Y P, Sun M, Chua K M, Wai L L, Liu D and Gaucher B P 2007 Antenna-in-package in LTCC for 60-GHz radio *IEEE Int. Workshop on Antenna Technology, March 21–23, 2007* pp 279–82

[6] Z. Qi, and B. Liang, Design of microstrip antenna with broader bandwidth and beam. *IEEE Antennas and Propagation Society International Symposium* 3A, 617–620 (2005) July.

[7] Shaoyong Wang[1], Qi Zhu[1] and Shanjia Xu[1]*Dept. of EEIS, Univ. of Sci. and Tech. of China, Hefei, 230027, Anhui, China* Received 28 January 2007 Accepted: 17 April 2007 Published online: 10 May 2007

[8]. David M. Pozar, "Consideration of millimeter wave printed antenna." *IEEE Trans on Antennas and Propagation*, vol. AP-31, NO. 5, Sept 1983

[9]. J.R. James and A. Henderson, "High frequency behavior of microstrip open circuit termination." *IEE J .Microwaves. Opt. Acoust.*, vol 3, Sept. 1979.pp.205- 218.

[10]. Richards, W.F., "An improved theory for microstrip antennas and applications" *IEEE Trans. On Antenna and Propagation*, Vol. AP-29, 1981, pp. 38-46.

[11]. K. R. Carver and J. W. Mink, "Microstrip antenna technology," *IEEE Trans. Antennas Propagat.*, vol AP-29,pp 25-37, Jan. 1981

[12]. Dubost , G., "Transmission line model analysis of lossy rectangular patch,*" Electron.Lett.*, vol. 18,1982,pp-86.

[13]. M. Kara. "An efficient technique for the computation of the input resistance of rectangular microstrip antenna elements with thick substrates," *Microwave opt. Technol. Lett.*, Vol. 13, No.6, Dec. 1996,pp.363-368.

[14]. R.Garg, P.Bhartia, I. Bhal, and A. Ittipiboon, Microstrip Antenna Design Handbook, Artech House, 2001

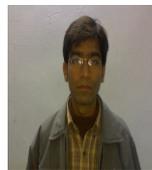

**Mr. Harsh kumar** received his B.Tech in 2008 from B N Mandal University, Madhepura, Bihar, India. Presently he is pursuing his M.E from Birla Institute of Technology Mesra, Ranchi, India, in the field of Microwave engineering. His areas of interest are Microstrip Antennas and RF circuit.

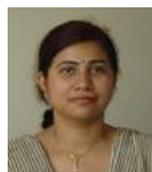

**Dr. Shweta Srivastava** received her Ph.D. from BHU, Varanasi (U.P.)-India. Presently she is working as senior lecturer in the Department Electronics & Communication Engg., B.I.T, Mesra, Ranchi, India. She is currently supervising a project named "Design of Active Microstrip Antenna for Wireless Communication" funded by DST, SERC Fast Track Scheme for Young Scientists. Her areas of interest are Microstrip Antennas, Electromagnec